\documentclass[published]{JHEP3} 

\JHEP{00(2008)000}

\JHEPspecialurl{http://jhep.sissa.it/JOURNAL/JHEP3.tar.gz}

\usepackage{epsfig,multicol,bbm}

\def\bequ{\begin{equation}}
\def\eequ{\end{equation}}
\def\barr{\begin{array}}
\def\earr{\end{array}}
\newcommand\fverb{\setbox\fverbbox=\hbox\bgroup\verb}
\newcommand\fverbdo{\egroup\medskip\noindent%
			\fbox{\unhbox\fverbbox}\ }
\newcommand\fverbit{\egroup\item[\fbox{\unhbox\fverbbox}]}
\newbox\fverbbox

\title{On the interaction between two Kerr black holes}

\author{Carlos A. R. Herdeiro, Carmen Rebelo\\
	Departamento de F\'\i sica e Centro de F\'\i sica do Porto, \\ Faculdade de Ci\^encias da
Universidade do Porto, \\ Rua do Campo Alegre, 687,  4169-007 Porto, Portugal\\
	E-mail: \email{crherdei@fc.up.pt,mrebelo@fc.up.pt}}

\received{September 3,2008} 		
\accepted{September 26, 2008}		

\preprint{0808.3941 [gr-qc]}	

\abstract{The double-Kerr solution is generated using both a B\"acklund transformation and the Belinskii-Zakharov inverse-scattering technique. We build a dictionary between the parametrisations naturally obtained in the two methods and show their equivalence.  We then focus on the asymptotically flat double-Kerr system obeying the axis condition which is $\mathbb{Z}_2^{\phi}$ invariant; for this system there is an exact formula for the force between the two black holes, in terms of their physical quantities and the coordinate distance. We then show that 1) the angular velocity of the two black holes decreases from the usual Kerr value at infinite distance to zero in the touching limit; 2) the extremal limit of the two black holes is given by $|J|=cM^2$, where $c$ depends on the distance and varies from one to infinity as the distance decreases; 3) for sufficiently large angular momentum the temperature of the black holes attains a maximum at a certain finite coordinate distance. All of these results are interpreted in terms of the dragging effects of the system.}


\begin{document} 


\section{Introduction}
Any relativistic gravitational theory which reduces, in an appropriate limit, to Newtonian gravity, must contain gravitomagnetic effects \cite{schutz}. In particular these include forces between mass currents and therefore spin-spin forces between classical rotating (and therefore extended) objects. In General Relativity, the force on a spinning body at rest in the exterior field of an arbitrary, stationary, rotating source is given by, to leading order in the distance  \cite{Wald:1972sz},
\bequ
\vec{F}\simeq -\frac{G}{c^2}\nabla\left(\frac{-\vec{j}\cdot \vec{J}+3(\hat{r}\cdot \vec{j})(\hat{r}\cdot \vec{j})}{r^3}\right) \ \qquad \ \stackrel{\vec{j}=j\hat{r},\vec{J}=J\hat{r}}{\Rightarrow} \ \qquad \  \vec{F}\simeq \frac{G}{c^2}\frac{6jJ}{r^4}\hat{r} \ , \label{spinspinforce} \eequ
where $\vec{j},\vec{J}$ are the spins of the test body and the source of the gravitational field, respectively, and $\vec{r}=r\hat{r}$ is the separation vector. This result  was obtained by considering a spinning test particle \cite{Papapetrou:1951pa} in the linearised Kerr field. It is nicely interpreted as a gravitomagnetic dipole-dipole interaction, with exactly the same form, except for the coupling constant and the sign, as its counterpart in magnetostatics. For spins aligned with the direction of separation, the result (\ref{spinspinforce}) yields a repulsive/attractive force for parallel/anti-parallel spins, in agreement with general arguments \cite{schutz}.

A natural question is if a repulsive spin-spin force can balance the attractive gravitational force between two classical spinning sources. A naive estimate, based on two spinning sources with spin $J$, mass $M$ and (\ref{spinspinforce}) reveals only that, if this is to be achieved, their separation will be of the order of their minimum size, $J/Mc$. The question cannot be answered at the level of a test particle approximation, since such balance implies that the approximation breaks down \cite{Wald:1972sz}. To avoid the complexity of an exact solution, one can consider the interaction between two spinning particles by an approximation method: a solution to Einstein's equations up to quadratic order in the metric perturbation around flat space, i.e. a post-post Newtonian approximation. Imposing asymptotic flatness one finds the suggestive result for the force between the two spinning particles (with spins aligned with the direction of separation) \cite{Bonnor:2001}
\bequ
F\simeq -G\frac{M_1M_2}{\zeta^2}+\frac{G}{c^4}\frac{6J_1J_2}{\zeta^4} \ , \label{forcebonnor} \eequ
where $M_i,J_i$ are the masses and angular momenta of the two particles and $\zeta$ is the coordinate distance (in Weyl canonical coordinates) between them. This force, computed from the energy momentum tensor associated to the conical singularity at the axis between the two particles, is just the sum of the Newtonian and the spin-spin term that one might have naively expected, confirming the form (\ref{spinspinforce}) for the latter. However, (\ref{forcebonnor}) is only meaningful if there exists an axis, i.e a set of fixed points of the azimuthal Killing vector field, between the two spinning particles, at which the conical singularity is defined. The same setup reveals an \textit{axis condition}, $J_1/M_1+J_2/M_2=0$, which is incompatible with $F=0$ for positive mass particles. Thus, already at this level of approximation we can foresee that no balance can be achieved.  

One might, however, question the validity of the approximation that leads to (\ref{forcebonnor}). Firstly, because it is unphysical to consider \textit{pointlike} classical spinning particles. Secondly, because one might suspect that more terms, other than the quadratic ones in $M,J$ given in (\ref{forcebonnor}), might contribute at the same distance approximation. If one wishes to describe the interactions between two classical spinning objects by a regular  (on and outside an event horizon) \textit{exact} solution of General Relativity, one is led to consider a solution describing two Kerr black holes in equilibrium.  The double-Kerr solution \cite{Kramer:79} was originally obtained by a solution generating technique which explores internal symmetries of the Ernst equation \cite{ernst} governing stationary, axi-symmetric solutions of the vacuum Einstein equations - the B\"acklund transformation \cite{stephani}. Subsequent analysis (see \cite{manko2001} and references therein) has revealed that, for two under-extreme objects, the solution cannot be made free of singularities and hence force balance cannot be achieved. The force between the two Kerr black holes takes the form, in a large distance expansion, \cite{dietz}\footnote{This formula is obtained imposing the axis condition and asymptotic flatness. We should note that we have corrected a sign in the force formula given in \cite{dietz} (cf. (\ref{forceback}) below).}
\bequ
F\simeq -G\frac{M_1M_2}{\zeta^2}\left[1+\frac{G^2}{c^4}\left(\frac{M_1}{\zeta}+\frac{M_2}{\zeta}\right)^2-\frac{3}{c^4}\left(\frac{J_1}{M_1\zeta}+\frac{J_2}{M_2\zeta}\right)^2\right] \ , \label{forcedietz} \eequ
where $M_i,J_i$ are the Komar masses and angular momenta of the two black holes and $\zeta$ the coordinate distance. The leading order term in the product of the two Komar angular momenta is exactly of the form (\ref{spinspinforce}), including the numerical coefficient, and the only two quadratic terms in $M,J$ in this expression are exactly the ones in (\ref{forcebonnor}). However, as anticipated above, there are more contributions at the same order in the distance as the term of the form (\ref{spinspinforce}). These include the general relativistic corrections to the Newtonian force between two masses, already present in the double-Schwarzschild solution. But there are also two other terms depending on the spin. Moreover the overall contribution of the three leading spin terms is \textit{always} repulsive, which is apparently in contradiction with the intuition gained from (\ref{spinspinforce}), and \textit{minimised} if $J_1/M_1+J_2/M_2=0$. It should be noted that an exact, meaningful force formula, in terms of Komar masses and angular momenta, for the general double-Kerr solution seems extremely hard to obtain. Observe that, for such formula to be meaningful, one would need to impose, with full generality, the condition of vanishing NUT charge (which is generically present in the solution), besides the axis condition.

The double-Kerr solution can also be obtained using a different solution generating technique - the inverse scattering method \cite{Belinskii:78,Belinskii:79}. This technique has been greatly explored over the last few years in higher dimensional general relativity (see \cite{Emparan:2008eg} for a review). In particular, in \cite{Herdeiro:2008en} the method was used to generate a double Myers-Perry black hole, a five dimensional analogue of the double Kerr. The latter solution was generated using the inverse scattering method in \cite{Letelier:1998ft}. However, the conclusions of \cite{Letelier:1998ft} are qualitatively different from those of \cite{dietz}, namely that, for the physical case when the NUT charge vanishes, the spin-spin interaction is always repulsive but \textit{maximised} if $J_1/M_1+J_2/M_2=0$.

The first purpose of this paper is to clarify the apparent contradiction between the results obtained by the two different methods. In section 2 we will derive the Kerr solution by both the B\"acklund and inverse scattering method and build an explicit dictionary between the two, showing their consistency. We will conclude that (\ref{forcedietz}) holds and correct some statements in \cite{Letelier:1998ft}. 

\begin{figure}[h!]
\begin{picture}(0,0)(0,0)
\end{picture}
\centering\epsfig{file=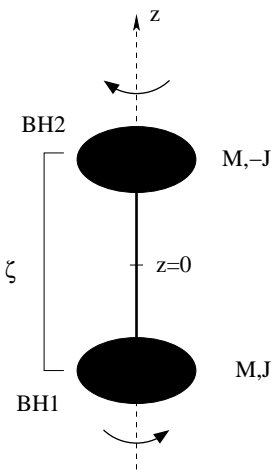,width=4cm}
\caption{Representation of the $\mathbb{Z}_2^{\phi}$ invariant double Kerr. This particular case of the double-Kerr solution is asymptotically flat and, for sufficiently low values of the angular momentum $J$ describes two equal mass and opposite angular momentum black holes with a strut (conical singularity) at the axis between them. This strut deforms the geometry of the horizon, deformation which we have made no attempt to represent in the figure.}
\label{z2symmetric}
\end{figure}

The second purpose of this paper is to consider the special case when the double Kerr system is $\mathbb{Z}_2^{\phi}$ invariant (cf. section 3) - figure \ref{z2symmetric}. The solution can then be reduced to a three parameter family (of two asymptotically flat Kerr black holes with an axis between them) from which some simple exact results can be obtained. It was already shown in  \cite{Varzugin:1998wf} that the \textit{exact} force between the black holes is given by
\bequ F=-G\frac{M^2}{\zeta^2}\left[1-\left(\frac{2MG}{c^2\zeta}\right)^2\right]^{-1} \ , \label{newforce} \eequ
where both black holes have Komar mass $M$ but opposite Komar angular momenta $J$ and $-J$; $\zeta$ is again the coordinate distance (in Weyl canonical coordinates). In this special case, the explicit angular momentum dependence vanishes from the force formula, which becomes identical to the one of the static case (see, for instance \cite{Costa:2000kf}). Herein, we will further show that:
\begin{description}
\item[1)] the exact angular velocity of the horizon of the two black holes is given by\footnote{Geometrised units, $G=1=c$ are used throughout this paper beyond this point.}
\bequ
\Omega_1=\frac{J}{2M\left(M^2+\sqrt{M^4-J^2\frac{\zeta-2M}{\zeta+2M}}\right)}\frac{\zeta-2M}{\zeta+2M}=-\Omega_2 \ , \label{omega12}\eequ
which varies from the Kerr value at $\zeta\rightarrow +\infty$ to zero at $\zeta\rightarrow 2M$ (corresponding to the two horizons touching) - figure \ref{omegatplot} (left);
\item[2)] The extremal limit of the black holes (i.e when they have zero temperature) depends on the coordinate distance, and is given by
\bequ
|J|=\sqrt{\frac{\zeta+2M}{\zeta-2M}} M^2 \ , \label{extremality} \eequ
which gives the usual Kerr ratio at $\zeta\rightarrow +\infty$ but tells us that, in the $\mathbb{Z}_2^{\phi}$ invariant double Kerr system, we can have a \textit{black hole with arbitrarily large intrinsic angular momentum for fixed mass}, by decreasing the coordinate distance;
\item[3)] The temperature of the black holes, for sufficiently high angular momentum, attains a maximum at a certain finite coordinate distance, cf. figure \ref{omegatplot} (right). 
\end{description}
Moreover, we will interpret all of these results, which will be presented in section 3, in terms of the mutual \textit{dragging of inertial frames} between the two black holes. We shall also offer a physical interpretation to the three spin interactions in (\ref{forcedietz}) which is consistent with the intuition gained from (\ref{spinspinforce}). 

We shall make some final remarks in section 4.

\section{The Double-Kerr solution}
The Double-Kerr solution is an exact, seven parameter family of vacuum solutions of four dimensional General Relativity, describing a stationary, axi-symmetric spacetime containing two Kerr black holes (or two naked singularities or one of each). Physically, the seven independent parameters can be thought to be 
\bequ  M_1,M_2,J_1,J_2,\zeta, b_{NUT}, M_{axis} \ ; \label{physicalparameters} \eequ
these are, respectively, the two Komar masses and the two Komar angular momenta of the individual black holes, the distance between them, the NUT charge of the spacetime and the Komar mass associated to the axis between the two black holes. The latter parameter is zero if the axis condition is obeyed. The existence of a Komar mass associated to the ``axis'' has been discussed in \cite{Manko:2004cq}, in the context of the double Kerr solution and in \cite{Herdeiro:2008en},  in the context of the double Myers-Perry solution.

We will now derive the solution using two different solution generating techniques. The full metric is extremely involved and we shall not present it explicitly,  although all the ingredients for its construction will be displayed. We will, however, give explicit expressions for the relevant physical quantities.

\subsection{B\"acklund Transformation}
\label{backlund}
A general stationary, axi-symmetric metric can be written in the form
\[ ds^2=-f[dt-\omega d\phi]^2+\frac{1}{f}\left[\rho^2d\phi^2+e^{2\gamma}(d\rho^2+dz^2)\right] \ . \]
All metric functions $f,\omega, \gamma$ depend on $\rho,z$ only. Solutions to the vacuum Einstein equations are determined by the Ernst potential $\mathcal{E}$, which we write in the form suggested by Yamazaki \cite{yamazaki}
\[ \mathcal{E}=\frac{E_-}{E_+} \ . \]
For the case of interest herein, $E_{\pm}$ are $2\times 2$ complex matrices defined as
\[E_{\pm}\equiv \det\left[\frac{n_i+n_k}{a_{ik}}\pm 1\right] \ , \qquad n_j\equiv \sigma_j e^{i\omega_j} \ , \qquad \sigma_j\equiv \sqrt{\rho^2+(z-a_j)^2} \ , \]
where $i=1,3$ ($k=2,4$) are line (column) indices and $a_{ik}\equiv a_i-a_k$. The solution is determined by the 3 real independent distances between the four points along the $z$ axis $a_1<a_2<a_3<a_4$ and the 4 real parameters $\omega_1,\omega_2,\omega_3,\omega_4$. The three metric functions are now determined as follows:
\[
f=\frac{A}{B} \ , \qquad A\equiv \Re(E_-E_+^*) \ , \qquad B\equiv E_+E_+^* \ ; \]
\[\omega=\frac{\Im(M^*E_+-LE_+^*)}{A}+k \ , \]
where $`*`$ denotes complex conjugation, $\Re$ ($\Im$) stands for real (imaginary) part and
\[L\equiv \sum_{t=1,3}\det\left[(n_t-n_k)\delta_{it}+\frac{n_i+n_k}{a_{ik}}(1-\delta_{it})\right] \ , \ \  M\equiv -\sum_{t=1,3}\det\left[-(a_{t0}+a_{k0})\delta_{it}+\frac{n_i+n_k}{a_{ik}}(1-\delta_{it})\right] \ , \]
$a_{j0}\equiv a_j-z$ and $k$ is a constant determined by the condition 
\[\lim_{\rho\rightarrow \infty}\omega f=0  \ , \]
meaning that $\phi=const$ is non-rotating at infinity (when the NUT charge vanishes);
\[ e^{2\gamma}=\bar{k}\frac{A}{ST} \ , \qquad S\equiv \det\left[\frac{2\sigma_i}{a_{ik}}\right] \ , \qquad T\equiv \det\left[\frac{2\sigma_k}{a_{ik}}\right] \ , \]
where $\bar{k}$ is a constant determined by the condition 
\[\lim_{\rho\rightarrow \infty} \frac{e^{2\gamma}}{f}=1 \ , \]
meaning that there are no conical singularities at infinity. 

Let us notice that instead of the seven parameters mentioned above it is more convenient to work with the following set
\bequ m_1,m_2,\lambda_1,\lambda_2,\alpha_1,\alpha_2,\zeta , \label{backparameters} \eequ
related to the previous set by
\[m_1=\frac{a_{21}}{2}\sec\frac{\omega_2-\omega_1}{2} \ , \qquad \zeta=a_{32}+\frac{a_{21}+a_{43}}{2} \ , \qquad m_2=\frac{a_{43}}{2}\sec\frac{\omega_4-\omega_3}{2} \ , \]
\[\alpha_1=\frac{\omega_1+\omega_2}{2} \ , \qquad \alpha_2=\frac{\omega_3+\omega_4}{2} \ , \qquad \lambda_1=\frac{\omega_2-\omega_1}{2} \ , \qquad \lambda_2=\frac{\omega_4-\omega_3}{2} \ . \]

We now consider the most relevant quantities to analyse the physics of this system. Consider  five regions denoted $I,II,III,IV,V$ according to $z<a_1$, $a_1<z<a_2$, $a_2<z<a_3$, $a_3<z<a_4$, $z>a_4$ and always $\rho=0$. 
\begin{itemize}
\item  The condition for asymptotic flatness is that 
\[ \omega|_{I}=2b_{NUT}=-\omega|_{V}=0 \ . \]
This means that the overall NUT charge, $b_{NUT}$, vanishes. Explicitly this yields
\bequ a_+^\alpha\zeta ^2 +2   m_1m_2 b_-\zeta+a_-^\alpha (m_2^2\cos^2\lambda_2 - m_1^2\cos^2\lambda_1) =0 \ .\label{flatback} \eequ
We have defined
\[
a_{\pm}^\alpha\equiv m_1\sin\alpha_1 \pm m_2\sin\alpha_2\ , \ \  b_{\pm}\equiv \cos\alpha_1\sin{\lambda_2}\pm \cos\alpha_2\sin{\lambda_1} \ , \ \  a_{\pm}^\lambda\equiv m_1\sin\lambda_1 \pm m_2\sin\lambda_2 \ . 
\]
\item  The axis condition, which guarantees that $\rho=0$ is a set of fixed points of the azimuthal Killing vector field, is 
\[ \omega|_{III}=0 \ . \]
Explicitly
\[
-[\zeta^2-m_1^2-m_2^2+(a_-^\lambda)^2+2m_1m_2\cos(\alpha_2-\alpha_1)][a_-^\alpha\zeta^2+2  m_1m_2b_+\zeta+a_+^\alpha(m_2^2 \cos^2\lambda_2-m_1^2\cos^2\lambda_1)] \]
\[+4 m_1m_2 \sin(\alpha_2-\alpha_1)[\zeta^2-m_1^2-m_2^2+(a_-^\lambda)^2]\zeta\]
\bequ
-4m_1m_2\cos(\alpha_2-\alpha_1) [a_+^\lambda\zeta^2+a_-^\lambda(m_2^2\cos^2\lambda_2-m_1^2\cos^2\lambda_1)]=0 \ .\label{axisback}\eequ

\item  The force between the two black holes is
\bequ F=\frac{1}{4}\left(1-e^{-\gamma}|_{III}\right)=-\frac{m_1m_2\cos{(\alpha_2-\alpha_1)}}{\zeta^2-m_1^2-m_2^2+(a_-^\lambda)^2-2m_1m_2\cos(\alpha_2-\alpha_1)}  \ . \label{forceback} \eequ
Note that, since this force is computed by the conical singularity in region III, it is only meaningful if the axis condition - which guarantees that there is an axis in region III, and hence that discussing a conical singularity is meaningful - is obeyed. Otherwise this is just a formal expression.
\end{itemize}

It is also relevant to note that the expressions for the Komar masses and angular momenta of the individual black holes take particularly simple (algebraic) forms in terms of the imaginary part of the Ernst potential $\Psi=\Im \mathcal{E}$ (the notation $\Psi|_{z=a_i}$ assumes implicitly that $\rho=0$):
\bequ M_1=-\frac{1}{4}\omega|_{II}\left(\Psi|_{z=a_2}-\Psi|_{z=a_1}\right) \ , \qquad M_2=-\frac{1}{4}\omega|_{IV}\left(\Psi|_{z=a_4}-\Psi|_{z=a_3}\right) \ , \label{massback} \eequ
\bequ J_1=\frac{1}{2}\omega|_{II}\left(M_1-\frac{a_{21}}{2}\right) \ , \qquad J_2=\frac{1}{2}\omega|_{IV}\left(M_2-\frac{a_{43}}{2}\right) \ . \label{amback}\eequ

\subsection{Inverse Scattering Method}
\label{inverse}
\begin{figure}[h!]
\begin{picture}(0,0)(0,0)
\put(75,46){$_{\left(1,-\frac{1}{q+\lambda_{II}}\right)}$}
\put(193,46){$_{\left(1,-\frac{1}{q+\lambda_{IV}}\right)}$}
\put(250,12){$_{(-(q+\lambda_{V}),1)}$}
\put(126,12){$_{(-(q+\lambda_{III}),1)}$}
\put(20,12){$_{(-(q+\lambda_{I}),1)}$}
\put(-5,33){$t$}
\put(-5,2){$\phi$}
\put(73,-8){$a_1$}
\put(111,-8){$a_2$}
\put(190,-8){$a_3$}
\put(228,-8){$a_4$}
\end{picture}
\centering\epsfig{file=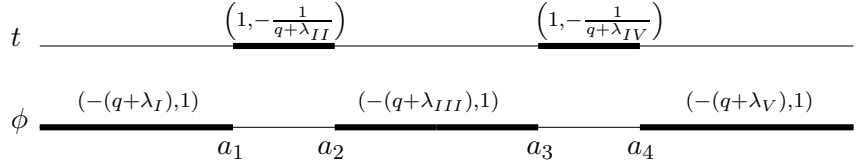,width=11cm}
\caption{Rod structure for the double Kerr spacetime. Next to each rod 
the corresponding eigenvector \cite{Harmark:2004rm} is displayed. For the double Schwarzschild solution these are simply $(1,0)$ for the timelike rods and $(0,1)$ for the spacelike ones.}
\label{doublekerr}
\end{figure}

In four spacetime dimensions, the inverse scattering method (or Belinskii-Zakharov method) \cite{Belinskii:78,Belinskii:79} can be used to construct new Ricci flat metrics with two commuting Killing vector fields from known ones,  by using purely algebraic manipulations. Such metrics can always be written in the form
\bequ
ds^2=G_{ab}(\rho,z)dx^adx^b+e^{2\nu(\rho,z)}(d\rho^2+dz^2) \ , \label{metricis}\eequ
where $a,b=1,2$. To use the BZ method, we need a seed solution, which we take to be the double-Schwarzschild solution \cite{bach,israelkhan}. It is defined by
\bequ 
G_0       = {\rm{diag}}  \left\{ -\frac{\mu_1 \mu_3}{\mu_2 \mu_4},\frac{\mu_2\mu_4 }{\mu_1\mu_3}\rho^2\right\} \ , \eequ
\begin{equation}
	e^{2\nu_0}=\frac{\kappa \, \mu_2 \mu_4 (\mu_1 \mu_2 + \rho^2)^2 (\mu_2 \mu_3 + \rho^2)^2 (\mu_1 \mu_4 + \rho^2)^2 (\mu_3 \mu_4 + \rho^2)^2}{\mu_1 \mu_3  (\mu_1 \mu_3 + \rho^2)^2 (\mu_2 \mu_4 + \rho^2)^2 (\mu_1^2 + \rho^2) (\mu_2^2 + \rho^2)(\mu_3^2 + \rho^2) (\mu_4^2 + \rho^2)} \ , \end{equation}
where $k$ is a constant to be determined below and 
\[{\mu}_k\equiv \sqrt{\rho^2+(z-a_k)^2}-(z-a_k) \ , \qquad \bar{\mu}_k\equiv -\sqrt{\rho^2+(z-a_k)^2}-(z-a_k) \ . \]
Its rod structure is represented in figure \ref{doublekerr}.

\bigskip

Using the method suggested by Pomeransky \cite{Pomeransky:2005sj}, we remove two anti-solitons, at $z=a_1,a_3$ and two solitons, at $z=a_2,a_4$. Thus, the seed metric becomes conformal to flat space
\begin{equation}
 G_0'       = \frac{\mu_2 \mu_4}{\mu_1 \mu_3}\, {\rm{diag}}  \left\{ -1,\rho^2 \right\}        \equiv  \frac{\mu_2 \mu_4}{\mu_1 \mu_3} \tilde{G}_0 \ .
\end{equation}
We will actually take the rescaled metric $\tilde{G}_0$ to be our seed (bearing in mind that one should multiply the final metric by the overall factor $\mu_2 \mu_4/\mu_1 \mu_3 $).  We take the generating matrix to be 
\begin{equation}
 \tilde{\Psi}_0(\lambda,\rho,z)= {\rm{diag}}  \left\{-1,\rho^2-2z\lambda-\lambda^2 \right\}  \ .
\end{equation}
One can verify that this matrix solves the Lax pair constructed in the
BZ method (see \cite{Belinskii:78,Belinskii:79}).  The double Kerr black hole is now obtained by a 4-soliton transformation:
using $\tilde{G}_0$ as seed, we add:
\begin{itemize}
\item one anti-soliton, at $z=a_1$ with
BZ vector $m_{0b}^{(1)}=(1,2a_1b_1)$; 
\item one anti-soliton, at $z=a_3$ with
BZ vector $m_{0b}^{(3)}=(1,2a_3b_3)$; 
\item one soliton at $z=a_2$ with
BZ vector $m_{0b}^{(2)}=(1,2a_2c_2)$; 
\item one soliton at $z=a_4$ with
BZ vector $m_{0b}^{(4)}=(1,2a_4c_4)$. 
\end{itemize}
Notice that we have introduced four new parameters: $b_1,b_3,c_2,c_4$.

Following the BZ algorithm one finds the double-Kerr solution.  The $t,\phi$ metric is given by
\[
G=\frac{\mu_2\mu_4}{\mu_1\mu_3}\tilde{G} \ , \qquad \tilde{G}_{ab}=(\tilde{G}_0)_{ab}-\sum_{k,l=1}^4\frac{(\tilde{G}_0)_{ac} m_c^{(k)}\left(\tilde{\Gamma}^{-1}\right)_{kl} m_d^{(l)}(\tilde{G}_0)_{db}}{\tilde{\mu}_{k}\tilde{\mu}_{l}} \ ,\]
where  $\tilde{\mu}_{k}=\mu_{k}$ for $k=2,4$ whereas $\tilde{\mu}_{k}=\bar{\mu}_{k}$ for $k=1,3$. The four vectors $m^{(k)}$ have spacetime components:
\bequ
m^{(k)}_a=m_{0b}^{(k)}\left[\tilde{\Psi}_0^{-1}(\tilde{\mu}_{k'},\rho,z)\right]_{ba} \ , 
\eequ
and the four BZ vectors where given above. The symmetric matrix $\tilde{\Gamma}$, whose inverse is $\tilde{\Gamma}^{-1}$,  is given by
\bequ
\tilde{\Gamma}_{kl}=\frac{m_a^{(k)}(\tilde{G}_0)_{ab}m_b^{(l)}}{\rho^2+\tilde{\mu}_{k}\tilde{\mu}_{l}} \ . \label{gamma} \eequ
The final quantity we need to have the complete metric is $\nu$, which is given by
\bequ
e^{2\nu}=e^{2\nu_0}\frac{\det{\Gamma_{kl}}}{\det{\Gamma^{(0)}_{kl}}} \ , \eequ
where $\Gamma^{(0)}$ and $\Gamma$ are constructed as in (\ref{gamma}) using $G_0$ and $G$, respectively.

The rod structure of the double-Kerr solution, depicted in figure \ref{doublekerr}, is completely defined by the quantities
\[ \lambda_i=\frac{G_{t\phi}}{G_{tt}}\Big|_{a_{i-1}<z<a_i} \ , \]
where $a_{0}\equiv -\infty$, $a_5\equiv +\infty$. Note that the metric $G$ is the one obtained directly from the BZ algorithm. Explicitly

\[\lambda _I=- 2\frac{(a_{21} a_{41} b_1+a_{32} a_{43} b_3)a_{42} + (a_{41}a_{43} c_2 + a_{21} a_{32}c_4)a_{31} b_1b_3}{a_{31} a_{42}(1+b_1 c_2b_3 c_4)+a_{32} a_{41}(b_1
c_2+b_3 c_4)+a_{21} a_{43} (b_3 c_2+b_1 c_4)} \ ,\]

\[\lambda _{II} = 2 \frac{(a_{21} a_{41} -a_{32} a_{43}b_1 b_3)a_{42} + ( a_{41}a_{43} c_2 + a_{21}a_{32}
c_4)a_{31}b_3}{a_{31} a_{42}(b_1- b_3 c_2 c_4)-a_{32} a_{41} (c_2- b_1 b_3 c_4)+a_{21} a_{43}(b_1 b_3 c_2-c_4)} \ , \]

\[\lambda _{III} = 2\frac{(a_{21}a_{42} c_2-a_{31} a_{43} b_3)a_{41}  + (a_{21} a_{31}
c_4-  a_{42}a_{43} b_1)a_{32}b_3 c_2}{a_{32} a_{41}(1+b_1 c_2b_3 c_4)-a_{21} a_{43}(b_1 b_3+c_2 c_4)+a_{31} a_{42} (b_1 c_2+b_3 c_4) } \ , \]

\[\lambda _{IV} = 2\frac{(a_{31} a_{41} + a_{32} a_{42}b_1 c_2)a_{43} + (a_{41}a_{42}b_3 - a_{31}a_{32}
c_4)a_{21}c_2}{a_{21} a_{43}(b_1 - b_3 c_2 c_4)+a_{32} a_{41}( b_3- b_1 c_2 c_4)+a_{31} a_{42}(b_1 b_3 c_2-c_4)}\  , \]

\[\lambda _V = 2\frac{(a_{21}a_{32} c_2+ a_{41}a_{43} c_4)a_{31} +(a_{32} a_{43} b_1 + a_{21}a_{41} b_3)a_{42}
c_2 c_4}{ a_{31} a_{42}(1+b_1 c_2b_3 c_4)+a_{32} a_{41}(b_1
c_2+b_3 c_4)+a_{21} a_{43} (b_3 c_2+b_1 c_4)}\  . \]
\bigskip
Moreover, $q$ turns out to be simply expressed as\footnote{$q$ could be defined by the coordinate transformation $dt\rightarrow dt-q d\phi$ which ensures that the resulting metric obeys $\lim_{\rho\rightarrow \infty}G_{t\phi}=0$, when the NUT charge vanishes.} 
\[ q=-\frac{\lambda_I+\lambda_V}{2} \ . \]
These quantities define:
\begin{description}
\item[i)]  the angular velocity of the horizon of the first and second black holes:
\[ \Omega^H_1=-\frac{1}{q+\lambda_{II}} \ , \ \ \ \ \ \Omega^H_2=-\frac{1}{q+\lambda_{IV}} \ , \]
\item[ii)] the NUT charge of the spacetime $b_{NUT}$; in particular, asymptotic flatness holds iff
\bequ -(q+\lambda_I)=2b_{NUT}=q+\lambda_V=0 \ , \label{flatbz} \eequ
\item[iii)] the axis condition
\bequ q+\lambda_{III}=0 \ , \label{axisbz} \eequ
\item[iv)] the regularity condition; choosing a periodicity for $\phi$  such that the only potential conical singularity is at $\rho=0$, $a_2<z<a_3$, the regularity condition turns out to be\footnote{Choosing the periodicity of $\phi$ in this way amounts to choosing the constant $k$ such that $\lim_{\rho\rightarrow \infty} e^{2\nu}=1$. The calculation of the regularity condition amounts to computing the period of $\phi$ in region III, which can be done using the procedure described in \cite{Harmark:2004rm}. Again note that if the axis condition is not obeyed, the result is formal, with no meaningful interpretation.}  
\bequ denominator[q]=denominator[\lambda_{III}] \ . \label{regularitybz} \eequ
\end{description}

\subsection{Dictionary}
There are seven independent parameters in the double Kerr solution. The ones that naturally arise in the inverse scattering method are
\bequ a_{21},a_{32},a_{43},b_1,b_3,c_2,c_4 \ . \label{BZparameters} \eequ
These are related to the set that arose in the B\"acklund transformation (\ref{backparameters}) via
\[m_1=\frac{a_{21}}{2}\frac{\sqrt{(1+b_1^2)(1+c_2^2)}}{1+b_1c_2} \ , \ \ \ \ \ m_2=\frac{a_{43}}{2}\frac{\sqrt{(1+b_3^2)(1+c_4^2)}}{1+b_3c_4} \ , \]
\[\tan\lambda_1=\frac{c_2-b_1}{1+b_1c_2} \ , \ \ \ \ \   \tan\lambda_2=\frac{c_4-b_3}{1+b_3c_4} \ ,\]
\[\tan\alpha_1=\frac{c_2+b_1}{1-b_1c_2} \ , \ \ \ \ \ \tan\alpha_2=\frac{c_4+b_3}{1-b_3c_4} \ , \]
\[\zeta=a_{32}+\frac{a_{21}+a_{43}}{2} \ . \]
Inverting these relations we have
\[a_{21}=2m_1\cos{\lambda_1} \ , \qquad a_{32}=\zeta-m_1\cos{\lambda_1}-m_2\cos{\lambda_2} \ , \qquad a_{43}=2m_2\cos{\lambda_2} \ , \]
\[b_1=\tan\frac{\alpha_1-\lambda_1}{2} \ , \ \ \ c_2=\tan\frac{\alpha_1+\lambda_1}{2} \ , \qquad b_3=\tan\frac{\alpha_2-\lambda_2}{2} \ , \ \ \ c_4=\tan\frac{\alpha_2+\lambda_2}{2}\ . \]
It is now a matter of algebra to verify that, using this dictionary, (\ref{flatback}) and (\ref{axisback}) exactly match (\ref{flatbz}) and (\ref{axisbz}). Moreover (\ref{regularitybz}) is equivalent to $F=0$ in (\ref{forceback}), i.e. 
\[m_1m_2\cos(\alpha_2-\alpha_1)=0 \ .\]
The latter equation together with (\ref{flatback}) and (\ref{axisback}) can be verified to exactly match expressions $(4.17b)$, $(4.18b)$ and $(4.20)$ in \cite{dietz}. Thus the results obtained with the inverse scattering method and B\"acklund transformation are consistent, as they should.

We can now comment on the results of \cite{Letelier:1998ft} mentioned in the introduction. The discrepancy originates from the fact that the quantities identified therein as being the black holes mass and angular momentum are \textit{not} the Komar quantities of the black holes. Thus, although the force formula written in \cite{Letelier:1998ft} (formula (34) therein) is correct, and indeed equivalent to formula (\ref{forceback}) above, its interpretation is not.\footnote{We can easily build a dictionary between the parametrisation used in \cite{Letelier:1998ft}, which we denote $\bar{a}_i,\bar{b}_i,\bar{m}_i,\bar{z}_i$, and the parametrisation used in section \ref{backlund}:
\bequ \bar{m}_i=m_i\cos\alpha_i \ , \qquad  \bar{b_i}=m_i\sin\alpha_i \ , \qquad \bar{a}_i=m_i\sin\lambda_i \ , \qquad \bar{z_2}-\bar{z}_i=\zeta \ . \label{dictionary2} \eequ} Moreover, the force formula is only meaningful, as we have emphasised, when both the axis condition and asymptotic flatness are guaranteed. But the condition imposed in  \cite{Letelier:1998ft} to guarantee vanishing NUT charge ($\bar{b}_i=0$) does not guarantee asymptotic flatness, as follows straightforwardly from (\ref{dictionary2}) and (\ref{flatback}).

\subsection{Physical Quantities}
The calculation of the Komar masses and angular momenta of the individual black holes can actually be simplified by using \textit{both} methods. Noting that
\[ \omega_{II}=-(q+\lambda_{II}) \ , \qquad \omega_{IV}=-(q+\lambda_{IV}) \ , \]
then (\ref{massback}) and (\ref{amback}) become
\bequ M_{1}= \frac{q+\lambda_{II}}{4} ( \Psi|_{z=a_2}-\Psi|_{z=a_1}) \ , \qquad M_{2}= \frac{q+\lambda_{IV}}{4} ( \Psi|_{z=a_4}-\Psi|_{z=a_3}) \ ,  \label{komarmasses} \eequ
\bequ J_1= -\frac{q+\lambda_{II}}{2} \left[ M_{1}- \frac{a_{21}}{2}\right] \ , \qquad J_2= -\frac{q+\lambda_{IV}}{2} \left[ M_{2}- \frac{a_{43}}{2} \right] \ . \label{komarangularmomenta} \eequ
The point is that whereas the $\Psi$'s are simpler to compute using the B\"acklund method, the $(q+\lambda)$'s are simpler to compute using the inverse scattering method. The former are  

\[
	\Psi|_{z=a_1}= \frac{a_{32}c_2 + a_{43} c_4 + a_{42}b_3 c_2 c_4}{a_{42} + a_{43} b_3 c_2 + a_{32}b_3 c_4}  \ , \qquad
	\Psi|_{z=a_2}= \frac{a_{31}b_1 + a_{43} c_4 + a_{41}b_1 b_3 c_4}{a_{41} + a_{43} b_1 b_3 + a_{31}b_3 c_4} \ , \]
\[	\Psi|_{z=a_3}= \frac{a_{21}b_1 + a_{42} c_4 + a_{41}b_1 c_2 c_4}{a_{41} + a_{42} b_1 c_2 + a_{21}c_2 c_4} \ , \qquad
	\Psi|_{z=a_4}= \frac{a_{21}b_2 + a_{32} b_3 + a_{31}b_1 b_3 c_2}{a_{31} + a_{32} b_1 c_2 + a_{21}b_3 c_2} \ . \]

We shall now observe a couple of consistency checks on the physical quantities computed. Firstly, for each black hole a Smarr formula should hold
\bequ
M_i=\frac{A_iT_i}{2}+2\Omega_i^HJ_i \ . \label{smarr} \eequ
The area of each black hole can be shown to be 
\[
	A_{1}= 2 \pi a_{21} (q+\lambda_{II})\frac{a_{31} a_{42}(b_1 - b_3 c_2 c_4) - a_{32} a_{41}(c_2 - b_1 b_3 c_4)-a_{21} a_{43}(c_4 - b_1 b_3 c_2)}{a_{31} a_{42}(1 + b_1 b_3 c_2 c_4)+ a_{32} a_{41}(b_1 c_2 + b_3 c_4)+a_{21} a_{43}(b_1 c_4 + b_3 c_2)} \ , \]
\[ A_{2}= 2 \pi a_{43} (q+\lambda_{IV})\frac{a_{21} a_{43}(b_1 - b_3 c_2 c_4) + a_{32} a_{41}(b_3 - b_1 c_2 c_4)-a_{31} a_{42}(c_4 - b_1 b_3 c_2)}{a_{31} a_{42}(1 + b_1 b_3 c_2 c_4)+ a_{32} a_{41}(b_1 c_2 + b_3 c_4)+a_{21} a_{43}(b_1 c_4 + b_3 c_2)} \ .
\]
Moreover, using the method described in \cite{Harmark:2004rm} to compute the temperatures, these can be shown to be very simply related to the areas:
\[ T_1=\frac{a_{21}}{A_1}  \ , \qquad T_2=\frac{a_{43}}{A_2} \ . \]
It follows, using also the results in section \ref{inverse} that the Smarr formula (\ref{smarr}) is obeyed for each black hole.

Secondly, we would like to check that, when $b_{NUT}=0$, the ADM mass/angular momentum is the sum of the Komar masses/angular momenta of the individual constituents of the spacetime. Note that we say ``constituents'' rather than black holes because, as pointed out in  \cite{Herdeiro:2008en}, when the axis condition is not obeyed, the rod in region III has a non-vanishing Komar mass and angular momentum. These are given by 
\bequ
M_{axis}= \frac{q+\lambda_{III}}{4} ( \Psi|_{z=a_3}-\Psi|_{z=a_2}) \ ,  \qquad J_{axis}= -\frac{q+\lambda_{III}}{2} \left[ M_{axis}- \frac{a_{32}}{2}\right] \ . \eequ
It can thus be verified that, for $b_{NUT}=0$, 
\[
\barr{l}
M_{ADM} = M_1+M_{axis}+M_2 \\ \\ 
\displaystyle{=\frac{ a_{31} a_{42} (a_{21} + a_{43})(1-b_1 b_3 c_2 c_4) + a_{32} a_{41} (a_{43}-a_{21})( b_1 c_2 - b_3 c_4) + a_{21} a_{43} (a_{41} + a_{32})( b_3 c_2 - b_1 c_4 )}{2(a_{31} a_{42}(1 + b_1 b_3 c_2 c_4) + a_{32} a_{41}( b_1 c_2 + b_3 c_4)+ a_{21} a_{43}( b_3 c_2 + b_1 c_4))}} \ , 
\earr \]
and a similar expression, albeit more involved, holds for $J_{ADM}$.

Let us conclude this section explaining how to obtain formula (\ref{forcedietz}) \cite{dietz}. One introduces two dimensionless parameters $\epsilon_i\equiv m_i/\zeta$. Then one imposes the axis and asymptotic flatness conditions to determine $\alpha_1,\alpha_2$ in terms of the other parameters to quadratic order in $\epsilon_i$. These expressions for the $\alpha_i$'s are then implemented in the formulae for the force, the Komar masses and angular momenta. The resulting expressions (to quadratic order) can be used to write the force in terms of the Komar quantities and the coordinate distance.

\section{The Double-Kerr solution with $\mathbb{Z}_2^{\phi}$ symmetry}
To interpret the exact force formula (\ref{forceback}) it should be written in terms of the physical quantities (\ref{physicalparameters}). Moreover, a meaningful interpretation would require that the conditions 
\bequ
M_{axis}=0=b_{NUT} \ , \label{axisNUT}\eequ
are obeyed. Although it does not seem possible to solve these conditions with full generality \cite{dietz}, there is a special case of particular interest: when the system has a $\mathbb{Z}_2^{\phi}$ invariance. By this we mean a variation of the usual $\mathbb{Z}_2$ invariance ($z\rightarrow -z$) that also reverts the $\phi$ direction: in the coordinates (\ref{metricis}) the $\mathbb{Z}_2^{\phi}$ is a discrete transformation that acts as 
\bequ z\rightarrow -z \ , \qquad \phi\rightarrow -\phi \ . \label{invariance}\eequ
Physically, invariance under (\ref{invariance}), means that the system is invariant under a $\pi$ rotation along an axis orthogonal to the $z$ axis and which intersects the middle point of the strut between the two black holes.  To realise this symmetry in the double-Kerr system, we require (\ref{axisNUT}) and, in terms of the Komar masses and angular momenta of the black holes
\bequ 
M_1=M_2\equiv M \ , \qquad J_1=-J_2\equiv J \ . \label{z2} \eequ
The resulting system has three independent quantities $M,J,\zeta$. Its ADM mass and angular momentum are $M_{ADM}=2M$, $J_{ADM}=0$. Although the spacetime has no overall angular momentum its individual constituents have.

\subsection{Physical Quantities}
In terms of the BZ parameters (\ref{BZparameters}) the constraints (\ref{axisNUT}) and (\ref{z2}) are implemented by the intuitive conditions
\[ a_{21}=a_{43}\equiv a \ , \qquad b_1=c_4\equiv b \ , \qquad c_2=b_3\equiv c \ . \]
The axis condition is immediately obeyed with no further restrictions. Asymptotic flatness determines $a$ as a function of $b,c,\zeta$
\[a=\frac{(c+b)(1+bc)}{(c-b)(1-bc)}\zeta \ . \]
The Komar mass and angular momentum can also be written in terms of $b,c,\zeta$ as 
\[M=\left(\frac{c+b}{c-b}\right)\frac{\zeta}{2} \ , \qquad J=\frac{c}{2(1-bc)}\left(\frac{c+b}{c-b}\right)^2\zeta^2 \ . \]
Note that the values of $b,c$ are restricted by the condition that $\zeta>a>0$ and $M>0$. The trivial case (no black holes) is $b=-c$. The static case (double-Schwarzschild) is obtained by taking
\bequ
b,c\rightarrow 0 \ , \ \ \ \ \ \frac{b}{c} \in ]-1,0[ \ ; \label{doubles}\eequ
it degenerates into flat space and a single Schwarzschild in the two limits of this interval, respectively. The extremal limit (zero temperature) is $1+bc=0$, which follows from the temperatures
\[T_1=\frac{1}{2\pi}\frac{(c-b)^2(1+bc)}{4c(c+b)\zeta}=T_2 \ . \]
 Note that the black holes have the same temperature. However they are \textit{not} in thermal equilibrium since they have a different chemical potential (symmetric angular velocities).\footnote{We would like to thank R. Emparan for this observation. A similar situation was discussed in \cite{Elvang:2007hg}.} Note also that the extremal limit does not correspond to $J=M^2$, as for a single Kerr black hole, but rather $J=cM^2$. Taking further $c=1$ (i.e. independent of $\zeta$) one ends up with the flat space limit. The angular velocity of the horizons are 
\[ \Omega_1^H=-\frac{b}{\zeta}\left(\frac{c-b}{c+b}\right)=-\Omega_2^H \ . \]
It can be verified that, in the allowed regions of the $b,c$ parameter space, $J_i$ has the same sign as $\Omega_i^H$, as expected. The limit $b\rightarrow 0$ ($c\neq 0$) corresponds to $a=\zeta$, i.e. the two horizons touch. In this limit the force
\[ F=\frac{(c+b)^2}{16bc} \ , \]
diverges and the angular velocity of the horizons vanishes. We shall interpret the latter conclusion in the next section.

Re-expressing the last three formulas in terms of $M,J,\zeta$, we obtain 
\[ F=-\frac{M^2}{\zeta^2}\left[1-\left(\frac{2M}{\zeta}\right)^2\right]^{-1} \ , \]
which is (\ref{newforce}) and 
\bequ
\Omega_{1} =\frac{2M - a}{4J}=-\Omega_2 \ , \qquad 	T_{1} =\frac{\zeta}{8\pi}\frac{M a (2M -a)}{2 J^2 (\zeta -2M)}=T_2 \ , 
\label{omegaz2}\eequ
where
\bequ a = 2 \sqrt{M^2- \frac{J^2}{M^2}\frac{\zeta-2M}{\zeta+2M}} \  , \label{a} \eequ
which in particular yields (\ref{omega12}). Note that the extremal limit is $a\rightarrow 0$, which yields (\ref{extremality}), and only corresponds to $J^2=M^4$ for $\zeta \rightarrow \infty$. The area of the black holes is

\bequ
	A_1 =\frac{8\pi}{\zeta}\frac{2 J^2(\zeta - 2M)}{M (2M-a)}=A_2 \ . 
\label{areaz2}
\eequ
Note that, in the large distance limit $\zeta\rightarrow \infty$, 
\[ \Omega_1=\frac{J}{2M(M^2+\sqrt{M^4-J^2})}=-\Omega_2 \ , \qquad T_1=\frac{\sqrt{M^4-J^2}}{4\pi M\left(M^2+\sqrt{M^4-J^2}\right)} \ ,  \]
\[ A_1=8\pi\left(M^2+\sqrt{M^4-J^2}\right)=A_2 \ , \]
which are the standard quantities for a single Kerr black hole. One also observes that in the touching limit $\zeta\rightarrow 2M$ all quantities become independent of the angular momentum:
\[ \Omega_1,\Omega_2\rightarrow 0 \ , \qquad T_1=\frac{1}{16\pi M}=T_2 \ , \]
\[A_1=32\pi M^2=A_2 \ . \]
Let us remark that the touching limit becomes a \textit{coalescence} limit corresponding to a \textit{single} Schwarzschild black hole if one takes simultaneously the limit (\ref{doubles}). This follows from the fact that the rod structure displayed in figure \ref{doublekerr} becomes, in such case, the one of a single Schwarzschild black hole.

\subsection{Physical interpretation}
\begin{figure}[h!]
\centering\epsfig{file=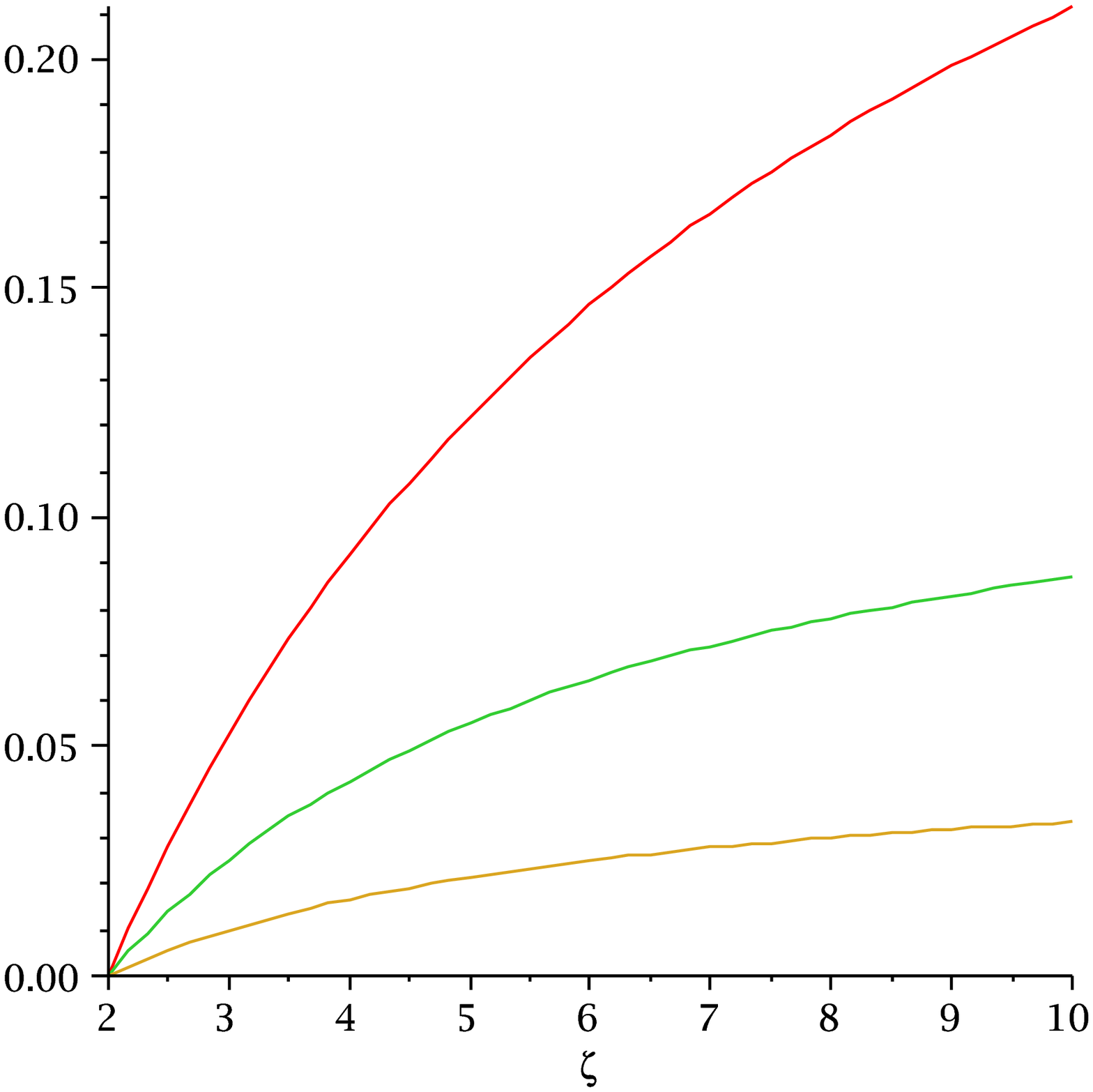,width=7cm}\ \ \ \ \
\centering\epsfig{file=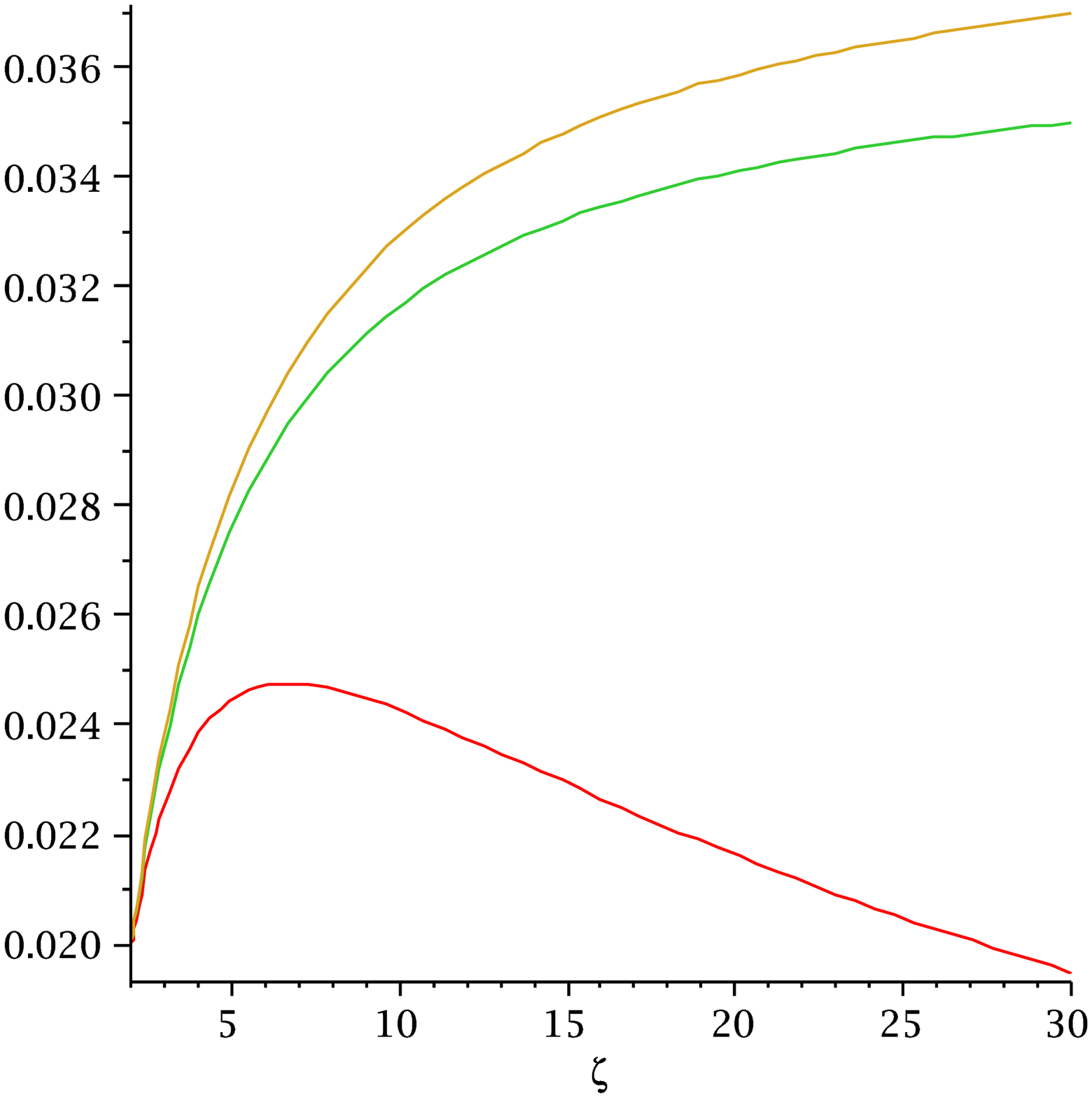,width=7cm}
\caption{Left (Right): $\Omega_1=-\Omega_2$ ($T_1=T_2$) for $M=1$ and various values of $J$, as a function of the coordinate distance $\zeta$.}
\begin{picture}(0,0)(0,0)
\put(-240,246){$\Omega_1=-\Omega_2$}
\put(-70,207){$_{J=1}$}
\put(-53,106){$_{J=0.2}$}
\put(-35,133){$_{J=0.5}$}
\put(-7,249){$T_1=T_2$}
\put(120,202){$_{J=0.5}$}
\put(100,122){$_{J=1}$}
\put(140,245){$_{J=0.2}$}
\end{picture}
\label{omegatplot}
\end{figure}

\begin{figure}[h!]
\centering\epsfig{file=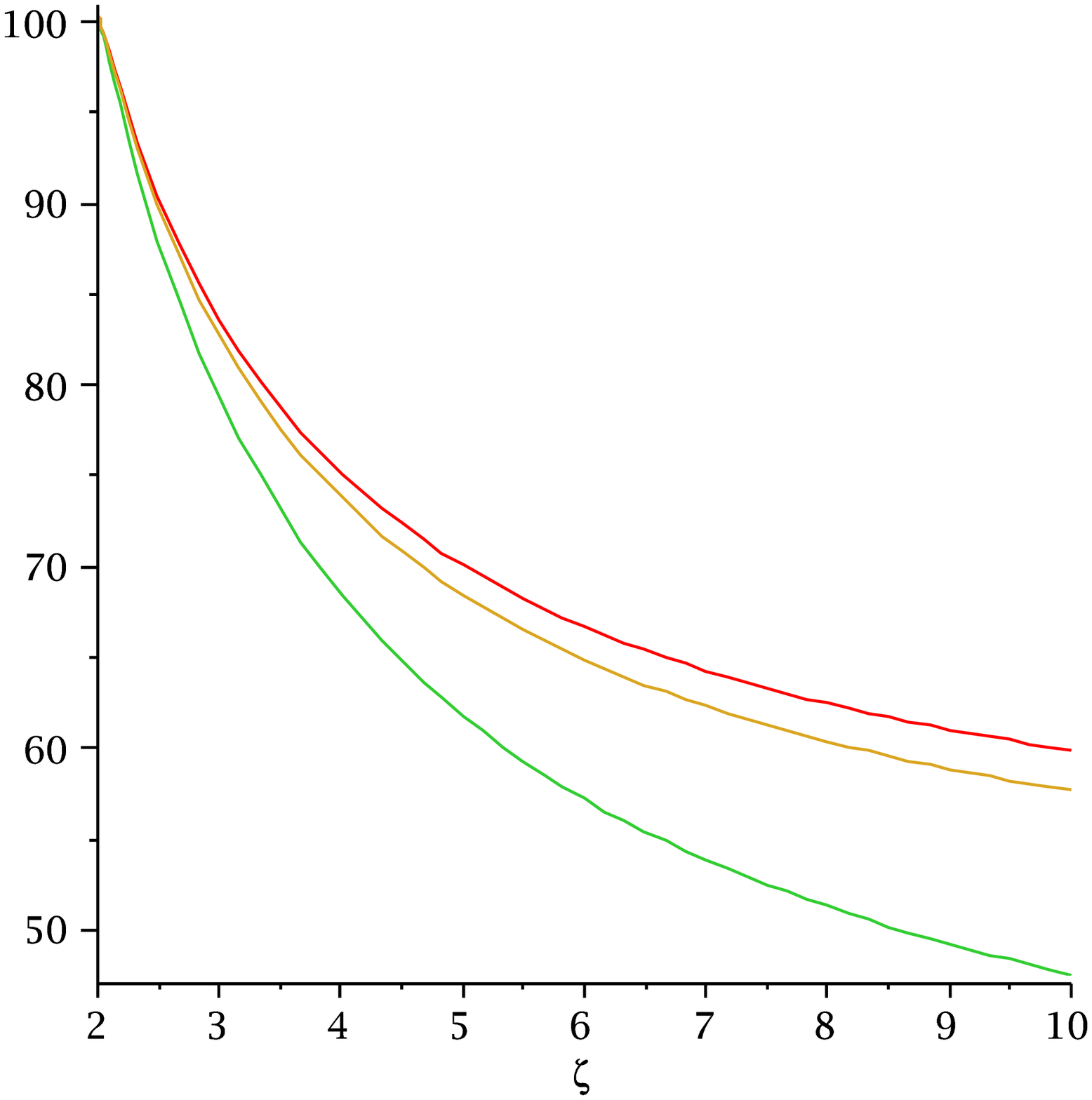,width=7cm}
\centering\epsfig{file=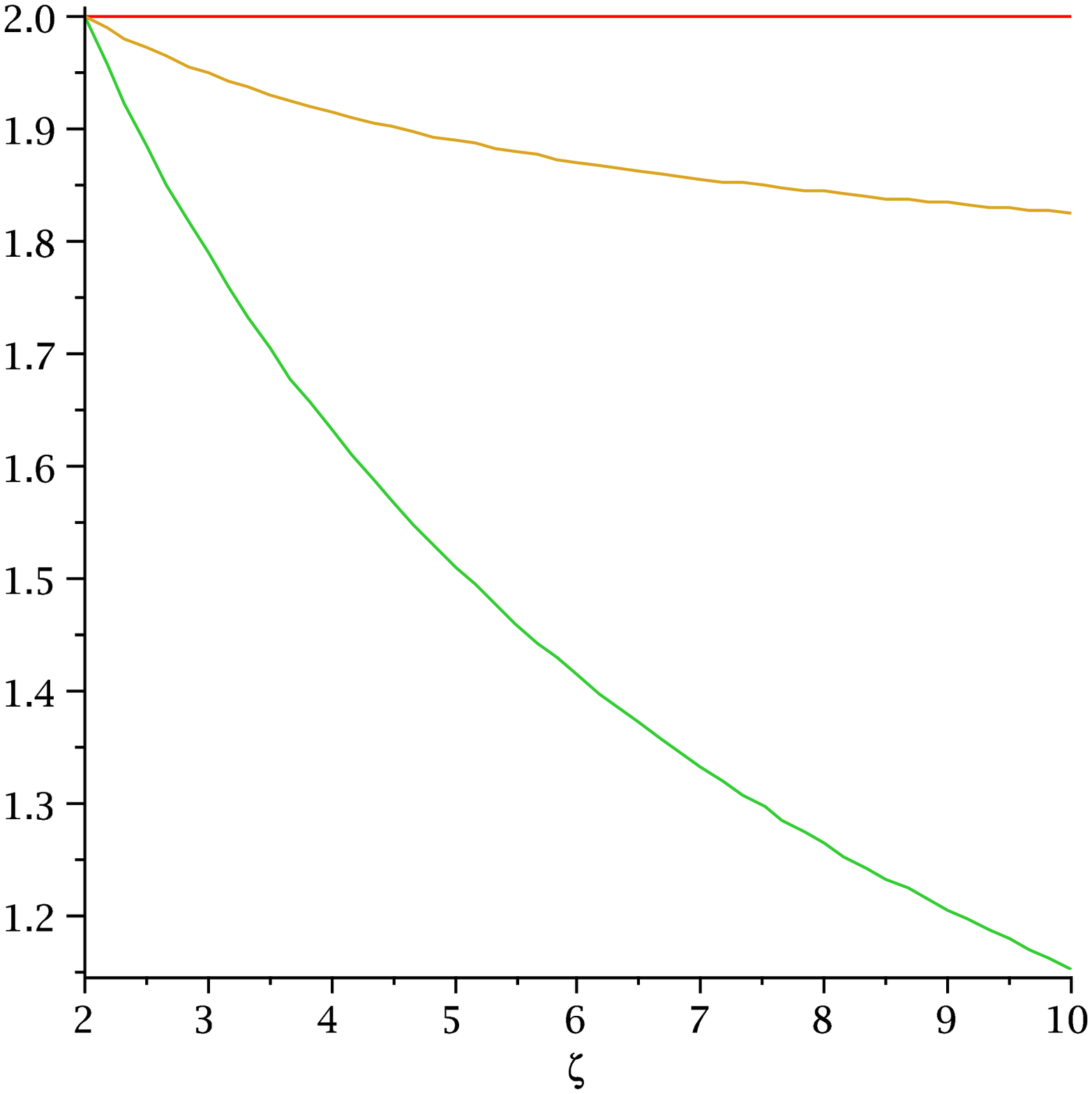,width=7cm}
\caption{Left: $A_1=A_2$ for $M=1$ and various values of $J$, as a function of the coordinate distance $\zeta$. Right: Parameter $a$ (3.6)  measuring deviation from extremality, as a function of $\zeta$.}
\label{areaplot}
\begin{picture}(0,0)(0,0)
\put(-240,243){$A_1=A_2$}
\put(-80,85){$_{J=1}$}
\put(-53,104){$_{J=0.5}$}
\put(-35,125){$_{J=0.2}$}
\put(-10,243){$a$}
\put(120,205){$_{J=0.5}$}
\put(100,141){$_{J=1}$}
\put(140,233){$_{J=0}$}
\end{picture}
\end{figure}

Let us start by analysing the angular velocity, temperature and area for the $\mathbb{Z}_2^{\phi}$ invariant double Kerr. These quantities are displayed in figures \ref{omegatplot} and \ref{areaplot}.

From figure \ref{omegatplot} (left) we see that, as we decrease the coordinate distance, the angular velocity of the two Kerr black holes decreases monotonically from the Kerr value (at $\zeta\rightarrow \infty$) to \textit{zero} in the touching limit ($\zeta\rightarrow 2M$). This can be attributed to the ``dragging of inertial frames'' from the other black hole. Take, say, black hole 1. It has a positive intrinsic angular momentum and hence a positive angular velocity. But it is dragged in the negative direction by the negative intrinsic angular momentum of black hole 2 and therefore it \textit{slows down}, i.e. its angular velocity is smaller than that of an isolated black hole with the same mass and angular momentum. And (\ref{omega12}) shows that the dragging effect  tends to \textit{cancel} the positive angular velocity due to the intrinsic angular momentum as the black holes touch. This could have been anticipated on physical grounds; a non-vanishing angular velocity in the touching limit would create a discontinuity in the spacetime dragging effects. Let us note that it had already been pointed out in \cite{Varzugin:1998wf} that in the $\mathbb{Z}_2^{\phi}$ double Kerr system the black holes would slow down each other, but without relating this fact to the dragging of inertial frames of General Relativity.

The variation of the temperature with the coordinate distance is, for sufficiently low angular momentum, essentially the same as in the static case (figure \ref{omegatplot} (right), $J=0.2,0.5$). In the static case, as we decrease $\zeta$ the black holes become larger (i.e the area increases monotonically, which also holds for the stationary case, cf. figure \ref{areaplot} (left)) and hence the temperature decreases monotonically. However, for sufficiently large angular momentum we see a new phenomenon occurring: there is a \textit{maximum temperature} at a certain, finite coordinate distance. We interpret the existence of such maximum as the result of two distinct effects. The first one is the one just discussed which also holds for the static case. The second one is intrinsically associated to rotation and to the fact that the extremality condition, $a=0$, depends on the coordinate distance, cf. (\ref{a}).

Recall that for a Kerr black hole, the temperature decreases monotonically from the Schwarzschild limit to the extremal limit. Hence, the temperature can be faced as a measure of the deviation from extremality. On the other hand, extremal rotating (uncharged) black holes are systems under the maximum allowed mechanical stress. A higher angular momentum (and hence higher angular velocity) would destabilise the horizon, transforming the system into a naked singularity. In the $\mathbb{Z}_2^{\phi}$ double Kerr system with fixed $M,J$ decreasing the coordinate distance \textit{slows down} the black holes, as discussed above, due to their mutual dragging effects. Thus, if the black holes were initially extremal (and hence in their mechanical limit) they relax and become non-extremal as they are brought closer - figure \ref{areaplot} (right) $J=1$. Since the temperature is a measure of the deviation from extremality, the temperature of the black holes should \textit{increase} as they are brought together, which is exactly what we can see in figure \ref{omegatplot} (right), $J=1$, for sufficiently large distance. As the black holes are brought sufficiently close, they slow down sufficiently so that the dominating effect is the same as in the static case, i.e. the increasing in size of the hole, and hence the temperature starts decreasing.

The fact that the extremality limit depends on $\zeta$ originates a novel phenomenon in four dimensional black hole physics. A Kerr black hole can have \textit{arbitrarily large angular momentum for fixed mass} in the $\mathbb{Z}_2^{\phi}$ invariant, asymptotically flat (and obeying the axis condition) double-Kerr system. This can be achieved simply by decreasing sufficiently $\zeta$ for the given value of the mass. Note, however, that these are not \textit{ultra-spinning} Kerr black holes, since they do not have an arbitrarily large angular velocity. It is well known that Myers-Perry black holes (with rotation in a single plane) \cite{Myers:1986un} have no upper bound for the angular momentum per unit mass in spacetime dimension $D\ge 6$. In the ultra-spinning limit they become ``pancake-like'' and it has been argued they should be unstable \cite{Emparan:2003sy}. In the present case, by contrast, since the angular velocity is always bounded and actually decreases with decreasing $\zeta$, for fixed mass and for the extremal solutions, we expect the horizon not to be too flattened by the rotation. It would, indeed be quite interesting to study in detail the horizon (and ergosphere) geometry of these solutions.\footnote{The fact that the extremality limit has $|J|>M^2$ has a counterpart for the charged dihole \cite{Chandrasekhar:1989ds,Emparan:1999au,Emparan:2001bb} wherein the extremal limit has $|Q|>M$. We thank R. Emparan for this observation.}

Let us now reconsider the generic double Kerr system, the force formula (\ref{forcedietz}) and, in particular, the three spin dependent terms. The first observation is that one of the $J^2_1,J^2_2$ terms will be present if a \textit{single} black hole has intrinsic angular momentum. Thus, unlike the $J_1J_2$ term it is not an interaction between the intrinsic rotation of both black holes. Secondly one notes that these $J^2_1,J^2_2$ terms are always repulsive. These two facts suggest that these terms are due to the interaction between the intrinsic spin of one black hole and the angular velocity that such intrinsic spin induces in the \textit{other} black hole, by dragging of inertial frames. Thus, the effect is present even if only one of the black holes has intrinsic spin and is repulsive, for the dragging originates an induced angular velocity parallel to the original spin. If this interpretation is correct one concludes that the spin-spin interaction being always repulsive means that the interaction between induced and intrinsic spins always puts an upper bound to the interaction between intrinsic spins.\footnote{Note that what is meant by `the interaction being always repulsive' is that turning on the angular momentum, for fixed \textit{coordinate distance}, the force between the black holes always decreases. The same does not apply for fixed \textit{physical distance}.}

\section{Final Remarks}
In this paper we have derived the double Kerr solution using two different solution generating techniques: the B\"acklund transformation and the inverse scattering method. We built a dictionary between the parametrisations that naturally arise in either method and used it to verify that some of the key quantities to analyse the solution - namely the asymptotic flatness condition, the axis condition and the force between the black holes - match. This allowed us to clarify an apparent contradiction between the conclusions of \cite{dietz} and \cite{Letelier:1998ft} concerning the spin-spin interaction between the two Kerr black holes. Both results are technically correct but the interpretation of the result given in the latter paper is not, as described in section 2.  

The analysis of the general solution led us to recognise one special case, which we have called the $\mathbb{Z}_2^{\phi}$ invariant double-Kerr, which simplifies sufficiently so that some exact statements can be made in terms of physically meaningful quantities, as discussed in section 3. Of particular interest are the physical consequences of the dragging of inertial frames. This is normally discussed in the context of a test particle approximation. It is quite striking to clearly see this dragging at the level of an \textit{exact} solution. One should mention this has primarily been achieved in the black Saturn solution \cite{Elvang:2007rd}, which has the advantage of being everywhere regular (on and outside the horizon), but the disadvantage of being a five dimensional solution and therefore perhaps less physical. 

Let us conclude with two questions. Firstly, is there a way to understand the area increase in the $\mathbb{Z}_2^{\phi}$ invariant system, for fixed mass and angular momentum, as the coordinate distance decreases? Note that this holds even for $J=0$ and is certainly a consequence of the horizon deformation due to the strut. Secondly, one could imagine the following ``gedankenexperiment'': set two Kerr black holes, with the same mass $M$ and counter-aligned spins, $J$ and $-J$, some distance apart $\zeta$. Let 
\[ \sqrt{\frac{\zeta+2M}{\zeta-2M}}>\frac{J}{M^2}>1 \ ; \]
thus, if the conclusions from the double Kerr system are applicable, the two Kerr black holes are under-extreme at their position but would be over-extreme at infinite distance. Could we give them initial velocities $-v$ and $v$ along the $z$ direction such that they reach a distance where they become \textit{over-extreme} and hence naked singularities? That would seem to violate both the second law of thermodynamics and cosmic censorship.

\section*{Note Added in the Revised Version}
After the pre-print of this paper appeared, another paper (arXiv:0809.2422 [gr-qc]) appeared considering the $\mathbb{Z}_2^{\phi}$ system. It correctly pointed out that the explicit metric written in our previous version (eq. (3.8) therein) did not describe the extremal limit of the $\mathbb{Z}_2^{\phi}$ system, asymptotically flat and obeying the axis condition. The corrected version of such metric is, however, much more involved, and therefore we chose not to include it in this revised version, but rather analyse it elsewhere. Note that this does not alter the correctness of any of the other results in this paper. Note also that the original physical statement that in the $\mathbb{Z}_2^{\phi}$ system one can have under-extreme black holes with $J>M^2$ was also stated in arXiv:0809.2422 [gr-qc] but, unfortunately, not acknowledged therein that it had been first discussed in our paper.

\section*{Acknowledgements}
We would like to thank L.F. Costa for discussions, M. Costa for comments on a draft of this paper and R. Emparan for correspondence. C.R. is funded by FCT through grant SFRH/BD/18502/2004. CFP is partially funded by FCT throught the POCI programme.

\end{document}